\documentclass{article}

\PassOptionsToPackage{numbers,sort,compress}{natbib}
\usepackage[preprint]{neurips_2023}
\usepackage[utf8]{inputenc}

\title{High-Resolution Convolutional Neural Networks on Homomorphically Encrypted Data via Sharding Ciphertexts}
\author{%
Vivian Maloney \quad Richard F. Obrecht \quad Vikram Saraph \quad Prathibha Rama \quad Kate Tallaksen \\
The Johns Hopkins University Applied Physics Laboratory\\
\texttt{\{vivian.maloney,freddy.obrecht,vikram.saraph\}@jhuapl.edu}\\
\texttt{\{prathibha.rama,kate.tallaksen\}@jhuapl.edu}\\
}

\usepackage{natbib}

\usepackage{fullpage}
\usepackage{graphicx}
\usepackage{subcaption}
\usepackage{amsmath}
\usepackage{amsfonts}
\usepackage{commath} 
\usepackage[colorlinks=true]{hyperref} 
\usepackage{xcolor} 
\usepackage{caption}
\usepackage{stfloats}
\usepackage{comment}
\usepackage{algorithm} 
\usepackage{algpseudocode} 
\usepackage{stmaryrd}
\usepackage{tabularx}
\usepackage{booktabs}       

\usepackage{amsmath}
\DeclareMathOperator{\sign}{sign}

\begin{document}

\maketitle

\begin{abstract}

Recently, Deep Convolutional Neural Networks (DCNNs) including the ResNet-20 architecture have been privately evaluated on encrypted, low-resolution data with the Residue-Number-System Cheon-Kim-Kim-Song (RNS-CKKS) homomorphic encryption scheme. We extend methods for evaluating DCNNs on images with larger dimensions and many channels, beyond what can be stored in single ciphertexts. Additionally, we simplify and improve the efficiency of the recently introduced multiplexed image format, demonstrating that homomorphic evaluation can work with standard, row-major matrix packing and results in encrypted inference time speedups by $4.6-6.5\times$. We also show how existing DCNN models can be regularized during the training process to further improve efficiency and accuracy. These techniques are applied to homomorphically evaluate a DCNN with high accuracy on the high-resolution ImageNet dataset, achieving $80.2\%$ top-1 accuracy. We also achieve an accuracy of homomorphically evaluated CNNs on the CIFAR-10 dataset of $98.3\%$. \footnote{After creating and uploading this manuscript, we became aware of related work published shortly before by Baruch et.\ al.\ \cite{IBM} }

\end{abstract}

\section{Introduction}

Deep learning has emerged as a powerful tool for solving image processing tasks due to its ability to automatically learn relevant features from raw data. Convolutional Neural Networks (CNNs), which are a type of deep learning model specifically designed for image processing, have achieved state-of-the-art performance on a variety of image processing tasks such as image classification~\cite{google_cat_paper}, object detection~\cite{yolo}, and segmentation~\cite{unet}.



Fully homomorphic encryption (FHE) ~\cite{gentry_og,first_he} is a technique enabling computation directly on encrypted data, and in particular, enabling Privacy Preserving Machine Learning (PPML). FHE has potential societal impact in applications where user and data privacy are critical, such as in cloud computing, healthcare analytics, and defense applications. However, adoption of FHE has been limited due to the speed of existing FHE neural network inference algorithms, and limitations of FHE itself. Previous work uses narrow or shallow DCNNs on low-resolution data, often using nonstandard activation functions, since FHE can only evaluate polynomials. Furthermore, it is challenging to ensure that polynomial approximations of activation functions are suitably accurate.

Key contributions of this work are summarized as follows:
\begin{itemize}
    \item We design and implement efficient homomorphic convolution and pooling algorithms, which have been parallelized and handle large inputs and channels via sharding techniques.
    \item We apply these algorithms to construct three families of ResNet architectures, achieving the highest homomophically evaluated accuracy on CIFAR-10 and ImageNet-1k while reducing the inference latency relative to the previous state-of-the-art. We also do not observe any degradation of encrypted model accuracy relative to its unencrypted counterpart.
    \item We propose a training technique to reduce the input range to our activation functions by penalizing the kurtosis of the distributions of BatchNorm outputs, allowing efficient homomorphic polynomial approximation of the GELU activation function.  
\end{itemize}

\section{Background}

\paragraph{Homomorphic encyption}

RNS-CKKS~\cite{rns-ckks, cheon2017homomorphic} is an FHE scheme that supports arithmetic over encrypted vectors of fixed-point numbers. Ciphertexts in this scheme are elements in the ring $R_Q^2$, where $R_Q = \mathbb{Z}_Q[x] / (x^{2N} + 1)$ and $Q$ is a large integer, and $2N$ is called the \emph{ring dimension}. Each such ciphertext has $N$ \emph{slots}, each of which stores a single real number, so it is useful to conceive of a ciphertext as a vector. Ciphertext vectors support vectorized addition and multiplication operations, as well as cyclic rotations. We pack images into RNS-CKKS ciphertexts.

Each ciphertext has a \emph{level}, or maximum number of multiplications that can be applied before decryption error becomes too high; each multiplication reduces the level by one. The ciphertext level is restored through \emph{bootstrapping}, though this is a time-consuming operation to be used sparingly.

\paragraph{Threat Model}

The threat model assumed is similar to previous PPMLs~\cite{lolaNets, lee2022low}. We encrypt the input image but not the model weights. A client homomorphically encrypts data it wishes to send, which is then sent to a server for processing. The server performs inference on the encrypted data directly, sending back the encrypted inference result to the client. Since it is assumed that only the client holds the secret key, only they can decrypt the result, which guarantees privacy from the server. Because the server does not see the decrypted inference result, the Li-Macciato attack~\cite{li2021security} is not applicable and we do not need to take noise flooding into account in our parameter selection.

\section{Related Work}

Early work on encrypted machine learning evaluated narrow and shallow CNNs with nonstandard activation functions on low-resolution data~\cite{gilad2016cryptonets, lolaNets}. Recent papers have begun evaluating larger CNNs with standard design features on encrypted data. Prior work on PPML most similar to ours are Multiplexed Parallel Convolutions~\cite{lee2022low} and TileTensors~\cite{aharoni2011helayers}. Multiplexed Parallel Convolutions homomorphically evaluates deep but narrow CNNs with standard activation functions on low-resolution data. TileTensors homomorphically evaluates shallow CNNs with nonstandard activation functions on high-resolution data. In this work, we homomorphically evaluate wide and deep CNNs with standard activation functions on high-resolution data. 

TileTensors uses concepts similar to our sharding approach to perform inference on $224\times224$ images using a modified AlexNet. They rely on shallow CNNs and do not perform the bootstrapping necessary to incorporate standard activation functions, instead relying on the same nonstandard activation function used in CryptoNets~\cite{gilad2016cryptonets} and LoLa Nets~\cite{lolaNets}, which is unsuited for DCNNs.

We improve on Multiplexed Parallel Convolutions, hereby defined as the multiplexed ResNet family, by supporting high-resolution images and wide channels that do not fit into a single ciphertext, as well as simplified packing. We also introduce a novel training regularization technique, enabling more efficient homomorphic evaluation of non-linear activations. Our implementation performs encrypted inference on a multiplexed ResNet-20 architecture $4.6\times$ faster than Ref.~\cite{lee2022low}. We homomorphically evaluate wide ResNet architectures not supported by the multiplexed algorithms, and achieve significantly higher accuracy than multiplexed architectures on standard datasets.

\section{Homomorphic Neural Network Operators}
\label{ops}

Algorithms have been carefully designed to minimize the number of encrypted multiplication and rotation operations to minimize latency. An \emph{image} consists of many \emph{channels}. All dimensions are assumed to be powers of two, and each channel is assumed to be square in shape. The approach is adaptable to dimensions not powers of two with appropriate rescaling or zero padding. Given an image with $c$ channels of size $m \times m$, we homomorphically encrypt and represent it with RNS-CKKS vectors. To encrypt an image into a ciphertext vector of size $m^2 c$, each channel $M^i$ is represented in row-major order, and they are concatenated to obtain a single plaintext vector.


\paragraph{Sharding and Encrypting an Image} \label{sect:shard-justification}

In RNS-CKKS, storage capacity of a single ciphertext is determined by the ring dimension of the scheme, and is typically in the range $2^{14}$ to $2^{16}$. If a $c \times m \times m$ tensor does not fit into a single ciphertext, channels are spread across \emph{multiple} ciphertexts, such that each ciphertext stores a subset of channels. Here, each ciphertext vector is called a \emph{shard}, and the maximum amount of data storable in a shard is called the \emph{shard size}. The performance of the scheme degrades with increasing ring dimension, so increasing the ring dimension to avoid sharding would negatively impact the efficiency of encrypted inference. 

\begin{figure}
  \centering
  \begin{subfigure}[c]{0.49\textwidth}
    \centering
    \includegraphics[width=\textwidth]{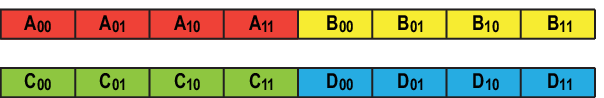}
    \caption{$4$ channels split across $2$ (image) shards.}
    \label{fig:image-shards}
  \end{subfigure}
  \begin{subfigure}[c]{0.49\textwidth}
    \centering
    \includegraphics[width=\textwidth]{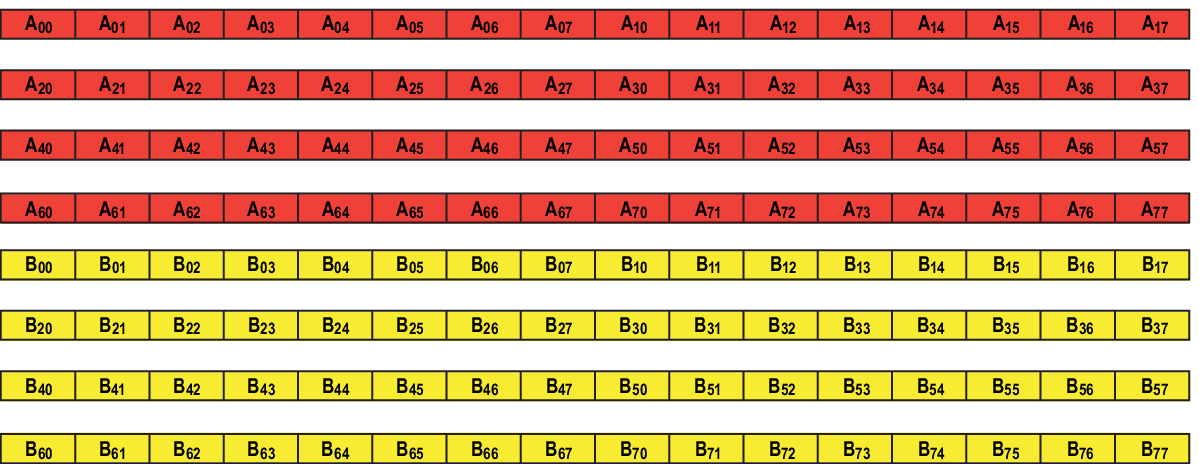}
    \caption{$2$ channels split across $8$ (channel) shards.}
    \label{fig:channel-shards}
  \end{subfigure}
  \caption{Illustrations of image sharding and channel sharding.}
  \label{fig:sharding}
\end{figure}

We distinguish the two cases of \emph{image shards} and \emph{channel shards}. For \emph{image shards}, a shard is large enough to hold at least one channel ($m^2 \leq s$), but multiple shards are needed to store all channels ($m^2c > s$). See Figure \ref{fig:image-shards} for an example of image shards. For \emph{channel shards} each channel must be split up across multiple shards ($m^2 > s$), so that each shard contains a set of consecutive rows from a single channel. See Figure \ref{fig:channel-shards}.

\paragraph{Duplicating and Permuting Channels}

If an image does not fill a shard, its channels are \emph{duplicated}. When $s > m^2c$, we define a \emph{duplication factor} given by $d = s / m^2c$, and place $d$ copies of each channel when concatenating them together. $d$ is tracked with the encrypted image as metadata. Our implementation of average pooling can \emph{permute} input channels. If one tracks the channels' order with a permutation defining the correct order, subsequent convolution operations can also be computed correctly. Therefore, we attach a channel permutation as metadata to an encrypted image.

\subsection{Convolution}
\label{subsect:conv}

\begin{figure}
\centering
\includegraphics[height=3in]{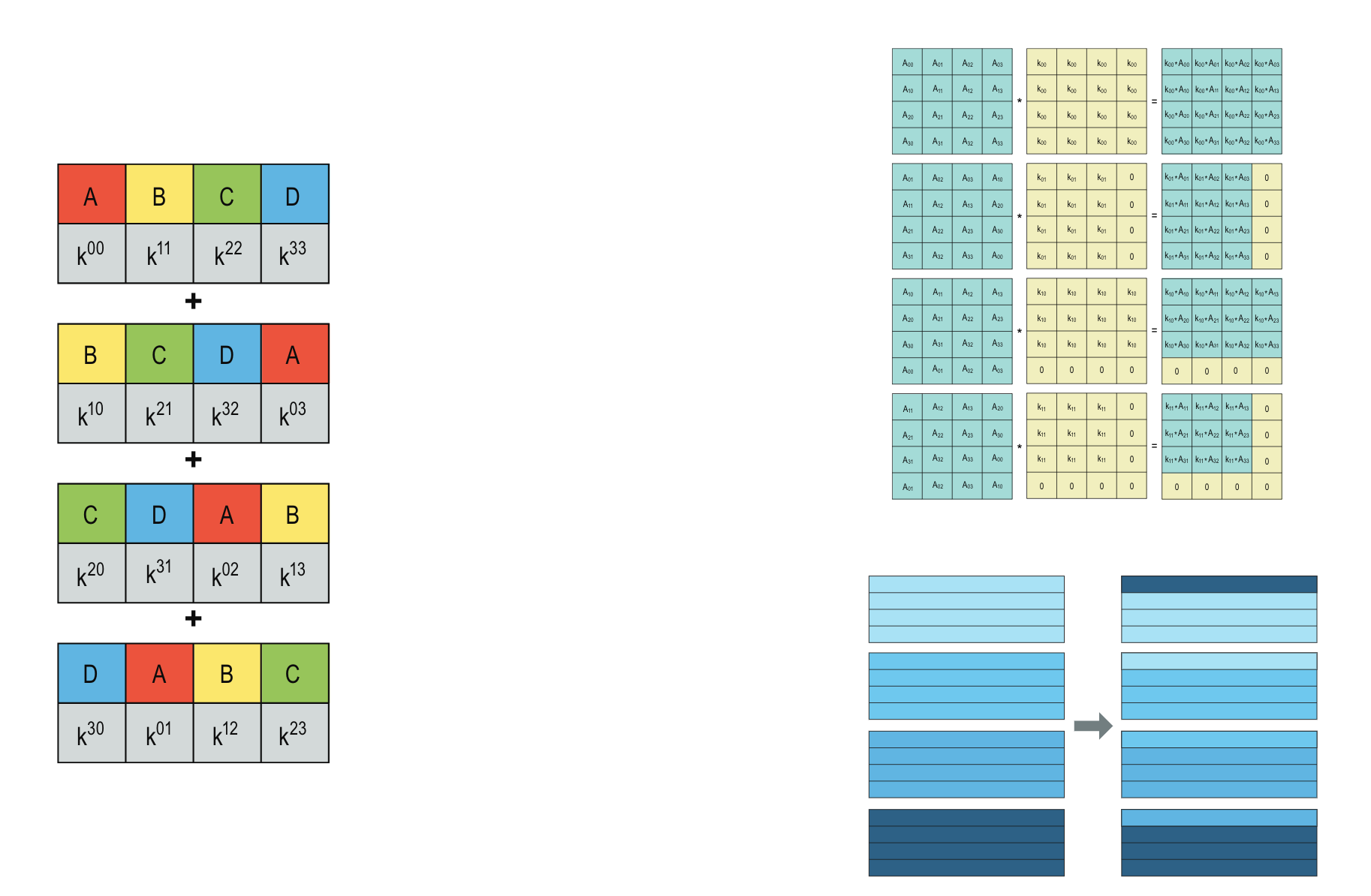}
%

\caption{(a) Partial convolution computation for a $4$-channel image convolved with a $1 \times 1$ kernel. (b) A single convolution computed by shifting the matrix. (c) Shifting rows from channel shards into adjacent ones.}
\label{fig:unified}
\end{figure}



We describe how to homomorphically convolve a single matrix with a single kernel, using same padding and a stride of $1$; this does not change the channel's dimensions. Convolution is typically thought of as sliding a kernel over as matrix. However, one may also think of convolution as fixing the kernel, and sliding the matrix, which is a more useful visual in what follows. We formalize this observation and use it to compute convolutions.
 Denote $\mathcal{S}_{k, \ell}$ on matrix $M$ as a function that shifts rows up by $k$ and columns left by $\ell$. $\mathcal{S}_{k, \ell}$ adds zeros when elements are shifted off the matrix. Then:
\begin{align}
    M * K = \displaystyle \sum_{k=-\kappa/2}^{\kappa/2} \sum_{\ell=-\kappa/2}^{\kappa/2} K_{k, \ell} \cdot \mathcal{S}_{k, \ell}(M).
\end{align}

See Figure \ref{fig:unified}. $\mathcal{S}_{k, \ell}$ is implemented homomorphically: shifting a row-major matrix by one column is done by rotating the ciphertext vector by $1$, while shifting by a row is done by rotating by $m$. Wrap-around elements are zeroed out by multiplying the ciphertext vector with an appropriate binary mask. This allows us to homomorphically compute $\mathcal{S}_{k, \ell}(M)$ for any shifts $k$ and $\ell$. To multiply $\mathcal{S}_{k, \ell}(M)$ by the scalar $K_{k, \ell}$, we create a vector of size $m^2$ and multiply $\mathcal{S}_{k, \ell}(M)$ elementwise with this vector. In practice, the multiplications for shift masking and those for kernel element multiplication are combined non-homomorphically before being applied homomorphically. 


\paragraph{With a Single Shard}
\label{subsubsect:single-conv}

Recall that to convolve a $c$-channel image with a single filter, $c$ matrix convolutions are individually computed, and the results are summed. An image is typically convolved with multiple filters to produce multiple channels. Convolutions are computed in parallel all at once.

Given an image $M$, denote $M^f_{ij}$ as the $(i, j)$-th element in the $f$-th channel of $M$. Filters $K$ ordinarily have dimensions $c_i \times c_o \times m \times m$, so that $K^{fg}_{ij}$ is $(i, j)$-th element in the kernel convolved with the $f$-th input channel used to compute the $g$-th output channel. We begin with a $1 \times 1$ kernel size, in which case $K^{fg}$ is the single-element kernel applied to the $f$-th input channel, to compute the $g$-th output channel. We further assume that $M$ fits in exactly one shard, and that $c_i = c_o = c$, so that $M * K$ also occupies one shard. Then the $g$-th channel of $M * K$ is given by Equation \ref{eqn:one-one}:

\begin{center}
\begin{minipage}{0.45\linewidth}
\begin{align}
(M * K)^g = \sum_{r=0}^{c-1} K^{r+g, g} \cdot M^{r+g} \label{eqn:one-one}
\end{align}
\end{minipage}
\hspace{0.5cm}
\begin{minipage}{0.45\linewidth}
\begin{align}
\bigparallel_{g = 0}^{c-1} K^{r+g, g} \cdot M^{r+g}
\label{eqn:concat}
\end{align}
\end{minipage}
\end{center}

where index arithmetic above is modulo $c$. We compute all $c$ output channels simultaneously. Given $0 \le r < c$, the $r$-th \emph{partial convolution} is defined in Equation \ref{eqn:concat}. The full convolution is obtained by summing over partial convolutions:

\begin{align} 
M*K &= \displaystyle \sum_{r=0}^{c-1} \bigparallel_{g = 0}^{c-1} K^{r+g, g} \cdot M^{r+g}
    = \displaystyle \sum_{r=0}^{c-1} \left ( \bigparallel_{g = 0}^{c-1}   K^{r+g, g} \cdot \bigparallel_{g = 0}^{c-1}  M^{r+g} \right ).
\end{align}

See Figure \ref{fig:unified} for a simple illustration of summing partial convolutions. Each summand corresponds to a single rotation of the ciphertext $M$ by $r \cdot m \cdot m$ positions.

When working with larger kernels, the prior approaches combine to compute the $g$-th output channel:
\begin{align}
    (M * K)^g = \displaystyle \sum_{r=0}^{c-1}
\sum_{k=-\kappa/2}^{\kappa/2} \sum_{\ell=-\kappa/2}^{\kappa/2} K_{k, \ell}^{r+g, g} \cdot \mathcal{S}_{k, \ell}(M^{r+g}).
\end{align}

Rotations $\mathcal{S}_{k, \ell}(M^{r+g})$ are computed once and cached. As with $1 \times 1$ kernels, we use partial convolutions to compute all $c$ channels at once.

Rather than directly implement strided convolution as in Ref.~\cite{lee2022low}, we instead compose an unstrided convolution with downsampling described in Section~\ref{sect:pooling}. This preserves the row-major order format and avoids multiplexed packing, and increases efficiency, as the multiplexed convolution algorithm of Ref.~\cite{lee2022low} has a multiplicative depth of 2, while we only use a single multiplicative level.

\paragraph{With Image Shards}

Let $M$ be an image of dimension $c_i \times m \times m$, split across $t$ shards, denoted as $[M]_0, \ldots, [M]_{t-1}$, implying a shard size $s = \frac{m^2 c_{i}}{t}$. Suppose we want to convolve $M$ with filters $K$ with dimensions $c_i \times c_o \times m \times m$. Then the $v$-th output shard, $[M * K]_{v}$, is computed as:
\begin{align}
    [M * K]_{v} = \displaystyle \sum_{u=0}^{t-1} [M]_u * K^{\iota(u), \iota(v)},
\end{align}

where $\iota(u)$ is the index interval $\iota(u) = [z \cdot u : z \cdot (u+1)]$, and $z = s / m^2$, or the number of channels per shard. Intuitively, each single convolution in the summand above is computed using the approach in the previous section~\ref{subsubsect:single-conv}, slicing $K$ accordingly, and summing up the results. With a shard size of $s$, $M * K$ is packed into $c_0 m^2 / s$ shards, and $v$ ranges over this.

\paragraph{Single Shard with Duplication and Permutation} \label{sect:duplication-and-permuted-channels}

Convolution must work with a shard with $d$-duplicated channels. Filters $K$ can be duplicated accordingly, but we instead index into $d$ times when computing $M * K$. Channels can also be permuted by pooling (see~\ref{sect:pooling}). In this case, the image passed from the previous layer is also assumed to return a permutation $\tau$ defining the correct channel order. To compute a convolution using this permutation, any time we were to index into the filter $K$ at input channel $i$ (so $K^i$), we instead index into $K$ at $\tau(i)$ (so $K^{\tau(i)})$.


\paragraph{With Channel Shards}


Convolving a channel-sharded image results in a channel-sharded image. Output channels are computed independently from one another, so we initially focus on convolving a shard of a single channel with a single kernel. Let $M^f$ be the $f$-th input channel of image $M$, which we convolve with a single kernel $K$. Let $[M^f]_u$ be the $u$-th shard. We cache all \emph{cyclic} rotations $\mathcal{S}_{k, \ell}([M^f]_u)$, for $k, \ell$ ranging over the indices of $K$. $[M^f * K]_v$ is computed from the cached rotations of the input shards.

Shifting channels requires shifting all associated shards simultaneously. Shifting columns is accomplished by shifting each shard independently. When shifting rows, one needs to shift rows of one shard into an adjacent shard. Each row shift is constructed from two cached rotations (with the exception of first and last shards). See Figure~\ref{fig:unified} showing how rows are shifted between shards.

Each output channel is computed by summing over row and column shifts, and each summand is itself a sum of two kernel-masked shards. That is:
\begin{align}
    [M^f * K]_v = \displaystyle \sum_{k=-\kappa/2}^{\kappa/2} \sum_{\ell=-\kappa/2}^{\kappa/2} \mathfrak{m}_{k, \ell}(K_{k, \ell}) \cdot \mathcal{S}_{k, \ell}([M^f]_v) + \overline{\mathfrak{m}_{k, \ell}}(K_{k, \ell}) \cdot \mathcal{S}_{k, \ell}([M^f]_{v + \sign{k}})
\end{align}

where $\mathfrak{m}_{k, \ell}(x)$ is the vector given by shard-size-many elements of all $x$, multiplied by the binary mask used in the shift operator $\mathcal{S}_{k, \ell}$, and $\overline{\mathfrak{m}_{k, \ell}}(x)$ is its complement. Then, to compute one shard $[M * K]_v$ of a single channel, we simply sum the shards $[M^f * K]_v$ over the input channels $f$. Each such shard is computed independently done in parallel, concluding channel-sharded convolution.

\subsection{Average Pooling} \label{sect:pooling}

We implement an average pooling operation with a $2 \times 2$ window; this increases the channel capacity of each shard by a factor of four. Our implementation preserves the format described previously, avoiding multiplexed packing used in Ref.~\cite{lee2022low}, which does not rearrange pixels after downsampling.

\paragraph{With Image Shards}

There are up to three steps involved with pooling: \emph{downsample}, which computes the average pool but leaves the original number of shards intact; \emph{consolidate}, which reduces the number of shards; and \emph{duplicate}, which duplicates channels if there is a single shard remaining.

\begin{figure}
  \centering
  \begin{subfigure}[t]{0.55\textwidth}
    \centering
    \includegraphics[width=\textwidth, scale=1.5]{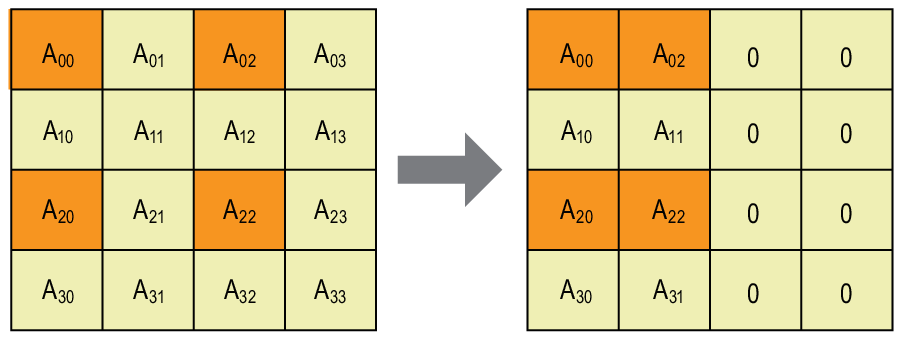}
    \caption{Horizontal reduction of a single $4 \times 4$ channel.}
    \label{fig:pool-horiz}
  \end{subfigure}
  \begin{subfigure}[t]{0.43\textwidth}
    \centering
    \includegraphics[width=\textwidth, scale=0.8]{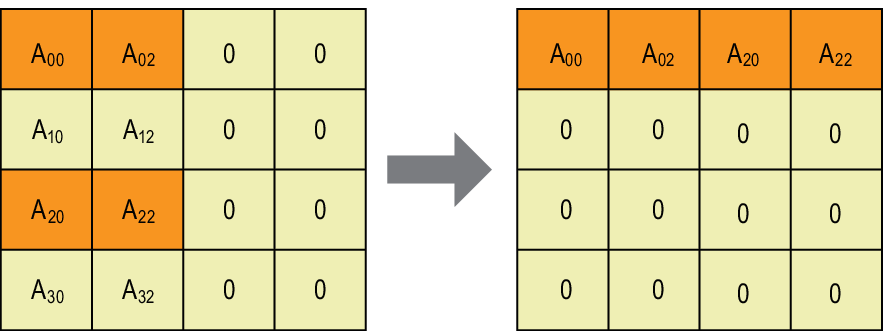}
    \caption{Vertical reduction of a single $4 \times 4$ channel.}
    \label{fig:pool-vert}
  \end{subfigure}
  \begin{subfigure}[t]{\textwidth}
    \centering
    \includegraphics[width=0.7\textwidth]{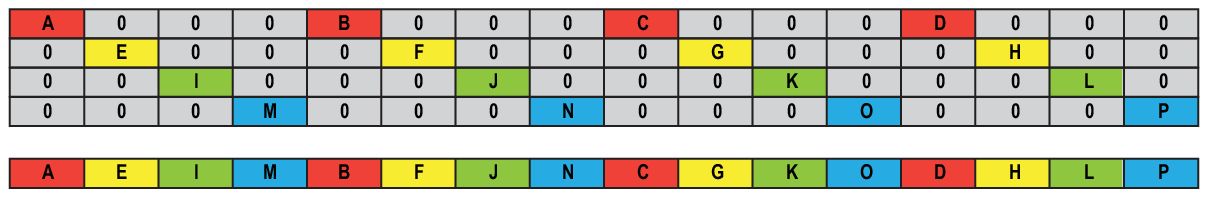}
    \caption{Consolidating four reduced shards into a single one. Each cell denotes a single channel; note that they are not in order, and are permuted, after consolidation.}
    \label{fig:pool-consolidate}
  \end{subfigure}
  \caption{Steps involved in a pooling operation. Duplication is not depicted.}
  \label{fig:pool}
\end{figure}


In the \emph{downsampling} step, we convolve each channel with a $2 \times 2$ kernel of $1$s (as we would with homomorphic convolutions). This replaces each $2 \times 2$ window in each channel with the sum of the elements in the window. Next, we want to select only one of four elements in the new $2 \times 2$ windows; we choose the top-left element. The following is how we operate on individual channels $M$, but generalizes to applying the operations to all channels within each shard simultaneously.

We \emph{horizontally reduce} the elements in channels of each shard, which is done with masking and summing over the channels of each shard, as in Equation \ref{eqn:horizontal}:

\begin{center}
\begin{minipage}{0.45\linewidth}
\begin{align}
M' = \sum_{i=0}^{(m-1) / 2} (M \cdot \mathfrak{m}_i) \ll i \label{eqn:horizontal}
\end{align}
\end{minipage}
\hspace{0.2cm}
\begin{minipage}{0.45\linewidth}
\begin{align}
\label{eqn:vertical}
M'' = \sum_{j=0}^{(m-1) / 2} (M' \cdot \mathfrak{m}_j) \ll 3i \cdot m / 2
\end{align}
\end{minipage}
\end{center}

where $\mathfrak{m}_i$ is the binary mask that selects elements in the $i$-th column of each channel $M$, and $\ll$ ($\gg$) denotes ciphertext rotation to the left (right) by $i$ slots. Then, we \emph{vertically reduce} each $M'$, as in Equation \ref{eqn:vertical}, where $\mathfrak{m}_j$ is the binary mask that selects the left half of $2j$-th row in $M'$. See Figures \ref{fig:pool}.


After downsampling, each $m \times m$ channel of the resulting shards contains only $m/2 \times m/2$ non-zero elements, all packed on the left-hand side. If we started with four or more shards, then we \emph{consolidate} the remaining non-zero elements into a quarter as many shards. This is done by 
rotating the shards from the previous step, and summing each group of four consecutive shards. 
\begin{align}
    S = S_0 + (S_1 \gg m^2) + (S_2 \gg 2m^2) + (S_2 \gg 3m^2).
\end{align}

Starting with two image shards, we only have two summands in the above, and with one shard, there is no consolidation step. Consolidating shards results in channels out-of-order. See Figure \ref{fig:pool-consolidate}. If we downsampled from two shards or fewer, then the resulting non-zero elements in the (single) consolidated shard would not fill up the entire shard, so we \emph{duplicate} the shard's channels:
\begin{align}
    S' = S + (S \gg m^2 / 4) + (S \gg 2m^2 / 4) + (S \gg 3m^2 / 4).
\end{align}

With two shards, we duplicate by a factor of two, so the above would only have two summands.

\paragraph{With Channel Shards}

Channel shards are downsampled individually, and every set of four consecutive shards is consolidated into one. In the edge case where the input image has one channel with two shards, we need to duplicate the resulting single shard by a factor of two. Pooling a channel-sharded image never results in an output with permuted channels.

\subsection{Other Layers}
\paragraph{Batch Normalization}

At inference time, batch normalization is an affine transformation, which is expressible as additions and multiplications, so can be implemented homomorphically. These are folded into kernel element multiplication and bias addition in the previous convolution, respectively.




\paragraph{Linear}

Evaluation of a linear layer is a matrix multiplication of the previous layer's output with the weights of the linear layer. Each element of a matrix multiplication is computed as a dot product. The dot product of one vector with another is computed by first taking their elementwise product, then summing the elements of the resulting vector. Elements of vector $v$ are summed by rotating over its slots, and adding the rotated vector to the original one. The result is a vector whose elements are all $\Sigma_i v_i$, and is done in logarithmically many rotations. We get a single activation in the linear layer's output, and repeat for each activation in the output of the linear layer. Activations are then masked and summed into a single ciphertext.

ResNets often pool each $m \times m$ input channel to a single pixel, and apply a linear layer at the end. In general, the pool could use a window size larger than $2 \times 2$, which we have not implemented directly. We fuse pool and linear into a \emph{pool-linear}. The linear layer's weights are duplicated as though it were operating on channels of size $m \times m$, and we divide by a normalization factor of $m^2$.


\paragraph{Gaussian Error Linear Unit (GELU)}

Non-linear activation functions are computed in RNS-CKKS through polynomial approximation. The polynomial degree and hence latency increases when the approximation must be accurate over a wide range. We introduce novel terms to the loss function during training to encourage hidden layer outputs to match the mean, variance, and kurtosis statistical moments of a Gaussian distribution, constraining the range over which the activation needs to be accurately computed. This allows more efficient low-degree polynomial approximation while minimally impacting model accuracy.


We use a GELU activation function since it is more amenable to polynomial approximations for fast homomorphic evaluation. We homomorphically compute a 59-degree polynomial approximation of GELU in a numerically stable way with a shallow arithmetic circuit by expanding the polynomial in a Chebyshev basis. Details on polynomial approximation of GELU and kurtosis regularization can be found in the Appendix.

\section{Empirical Results}
\label{results}

We use OpenFHE's implementation~\cite{OpenFHE} of FHE with RNS-CKKS to implement the neural network operators described in Section~\ref{ops} in C++, which are then thinly wrapped with Python bindings to build neural network architectures. Weights are loaded using PyTorch's API, though the approach is indepedent of deep learning framework. OpenMP is used to leverage parallelism from multicore CPUs. As our main focus is on fast encrypted inference of trained models rather than the unencrypted training process, we defer most of the details on the training configuration to the Appendix.

Experiments for ResNet-9 and multiplexed ResNets were run on a machine with a hyperthreaded AMD Ryzen Threadripper 3970X 32-core processor, 128 GB of memory, and an Ubuntu 22.04.2 operating system. Experiments for the encrypted ResNet-50 were run on a server with an AMD EPYC 7742 64-core processor, 800 GB of memory, and RHEL 7.9.

\subsection{Datasets}

We perform image classification on CIFAR-10, CIFAR-100, and ImageNet, using various ResNets to evaluate the performance of our homomorphic neural network operators. CIFAR-10 and -100 contain $32\times32$ color images in 10 and 100 classes, respectively~\cite{cifar}. ImageNet-1k is a much larger scale dataset containing over 1.2 million high-resolution images with 1000 different classes ~\cite{imagenet}, and is typically resized to $224\times224$ during inference, though this does not match our assumption that dimensions are powers of two. We evaluate two different models on ImageNet-1k resized to resolutions of both $128\times128$ and $256\times256$.

\subsection{Architectures}


We modify DCNN architectures to decrease encrypted inference latency without adversely affecting model accuracy. We use $2\times2$ average pooling with stride $(2,2)$, and the GELU activation function. We train models with kurtosis regularization as described in the previous section, and more extensively in the Appendix.\footnote{If using kurtosis-regularized GELU is not an option, such as when evaluating pre-existing models, our algorithms are compatible with any approach for computing ReLU over a wider range, such as higher-degree polynomial approximation or the approach in Ref.~\cite{lee2022optimization}.}.



We homomorphically evaluate three classes of ResNets on CIFAR-10 and -100. We first evaluate the narrow deep multiplexed ResNet family used in the previous state-of-the-art for homomorphic DCNNs Ref.~\cite{lee2022low}, as well as a wide ResNet-9 architecture taken from DAWNBench~\cite{coleman2017dawnbench}, and finally a fine-tuned version of the wide and deep ImageNet-1k ResNet-50 v1.5~\cite{he2016deep}. The wide ResNet-9 and -50 achieve substantially higher accuracy than the multiplexed family, achieving a best accuracy of 94.7\% and 98.3\% on CIFAR-10, respectively, surpassing the 92.95\% reported in Ref.~\cite{lee2022low} for a multiplexed ResNet-110 and the 92.8\% we achieved for a multiplexed ResNet-56.

Our ImageNet-1k architecture is modified ResNet-50 v1.5~\cite{he2016deep} with GELU and average pooling. This is a wide architecture, using between 64 and 2048 channels. On ImageNet-1k, we train and evaluate ResNet-50 on images resized to $128\times128$ and $256\times256$, respectively. The $256\times256$ model requires both channel shards and image shards, while the $128\times128$ model only requires image shards. As such, this illustrates a trade-off between model accuracy and inference time for image resolution. The resolution during training was set according to the FixRes~\cite{touvron2019fixing} optimization, where the training resolution is $3/4$ of evaluation resolution to account for data augmentation.


\begin{table}
    \caption{
    Model accuracy is averaged over five runs for all architectures except ResNet-50, and the quoted error is the standard deviation.
    The (*) represents our implementation of the multiplexed architectures found in Ref.~\cite{lee2022low}. Due to long training times, ResNet-50s are only trained once.}
    \centering
    \label{table:unencrypted_acc}
    \begin{tabular}{llcc}
    \toprule
    Dataset     & Model & Average Accuracy (\%) & Best Accuracy (\%)  \\ \midrule
    CIFAR-10    & ResNet-9    & $94.5 \pm 0.1$  &  94.7   \\
                & ResNet-50   & $98.3$          &  98.3   \\
                & ResNet-20*  & $90.6 \pm 0.3$  &  91.0   \\
                & ResNet-32*  & $92.2 \pm 0.2$  &  92.5   \\
                & ResNet-44*  & $92.2 \pm 0.1$  &  92.3   \\
                & ResNet-56*  & $92.8 \pm 0.2$  &  93.0   \\
                & ResNet-110* & $92.7 \pm 0.2$  &  92.8   \\ \midrule
    CIFAR-100   & ResNet-9    & $74.9 \pm 0.2$  &  75.3   \\
                & ResNet-32*  & $66.6 \pm 0.4$  &  67.0   \\ \midrule
    ImageNet-1k & ResNet-50 @ 128  & $74.1$     &  74.1    \\
                & ResNet-50 @ 256  & $80.2$     &  80.2   \\ \bottomrule
    \end{tabular}
\end{table}

\subsection{Encrypted Inference Discussion}

For the encrypted ResNet-50, we used a RNS-CKKS ring dimension of $2^{16}$ and shard size of $2^{15}$ with 59-bit scaling factors and a multiplicative depth of 34. When evaluating the multiplexed ResNets and ResNet-9, we used a lower shard size of $2^{14}$. This lower shard size trades slower initial layers for faster later layers and bootstrapping operations, and improved the encrypted latency for these narrower architectures. These parameters suffice for a standard 128-bit security level~\cite{albrecht2021homomorphic}.
The distributions prior to GELU are analyzed in order to determine a safe bound for our polynomial
approximations; see the Appendix for details. For each model, runtime experiments 
are collected for 25 inferences; for each run, the runtimes for each algorithm are 
summed, and then the average is displayed in Tables~\ref{table:multiplex_inference}
and~\ref{table:non_multiplex_results}, where the quoted error is the standard deviation in the total runtime. ResNet-9 and -50 models, which allow the channel dimension to substantially grow, spend less relative time bootstrapping when compared to the multiplexed ResNet family.


During inference on ImageNet-1k, ResNet-50 at 128 resolution uses a maximum of 32 shards, and at 256 resolution uses a maximum of 128 shards. On CIFAR-10, ResNet-50 uses a maximum of 16 shards. Due to channel size, inference on 256 resolution requires the use of channel shards, and has a $2.9\times$ slower latency. However, note that ResNet-50 on 256 resolution has a $6.1\%$ higher accuracy, so in this case, using higher resolution images produces a better classifier.

The \emph{logit residual}, which is the difference between decrypted and unencrypted logits,
generally form tight Gaussian distributions centered at zero. 
By using GELU and a small input range, we decreased the noise from bootstrapping and the polynomial approximation. This is reflected in the increased precision of the
logit residual distributions, which has standard deviations at the $10^{-4} - 10^{-2}$ level when fit to a Gaussian,
see Table 1 in the Appendix for more details. 
We ran 1000 inferences with ResNet-20 on CIFAR-10, and all encrypted predictions match respective
unencrypted predictions; this is an 
improvement over Ref.~\cite{lee2022low}, where the encrypted classification accuracy is $0.1-0.5$\% lower than the 
unencrypted accuracy. 
Furthermore, the difference in the top-2 logits between the encrypted and unencrypted ResNet-20 are examined, yielding Gaussian standard deviations at the $10^{-4}$ level.
Thus, using kurtosis and GELU allows us to perform faster and more reliable encrypted inference.


As further discussed in the Appendix, we determined that the logit error is mainly due to bootstrapping noise. By applying MetaBTS~\cite{bae2022meta} to reduce bootstrapping noise we further increased logit precision by a factor of $20\times$ at the expense of a $1.7\times$ increase in latency.

\begin{table}
    \caption{Average latency (seconds) for our implementation of the multiplexed ResNet architectures.}
    \centering
    \label{table:multiplex_inference}
    \begin{tabular}{lrrrrrr}
    \toprule
     & \multicolumn{5}{c}{CIFAR-10} & \multicolumn{1}{c}{CIFAR-100} 
    \\\cmidrule(lr){2-6}\cmidrule(lr){7-7}
               & \multicolumn{1}{c}{ResNet-20} & \multicolumn{1}{c}{ResNet-32}        &    \multicolumn{1}{c}{ResNet-44}  & \multicolumn{1}{c}{ResNet-56} & \multicolumn{1}{c}{ResNet-110} & \multicolumn{1}{c}{ResNet-32}\\\midrule
    ConvBN    & 116 & 182 & 247  & 314  & 610  & 180  \\
    GELU      & 49  & 66  & 84   & 101  & 180  & 65   \\ 
    Bootstrap & 309 & 432 & 558  & 681  & 1235 & 427  \\
    Avg Pool  & 9   & 9   & 9    & 9    & 9    & 9    \\ 
    Linear    & 3   & 3   & 3    & 3    & 3    & 23   \\ \midrule
    Our Total & $486 \pm 3$   & $692 \pm 5$ & $901 \pm 6$  & $1108 \pm 7$  
              & $2037 \pm 13$ & $704 \pm 5$  \\
              \midrule
    Total~\cite{lee2022low} & $2271$  & $3730$ & $5224$  & $6852$  
                            & $13282$ & $3942$  \\
      \bottomrule
    \end{tabular}
\end{table}

\begin{table}
    \centering
    \caption{Average latency (seconds) for our architectural contributions.}
    \label{table:non_multiplex_results}
    \begin{tabular}{lrrrrrr}
    \toprule
     & \multicolumn{2}{c}{CIFAR-10} & \multicolumn{1}{c}{CIFAR-100} & \multicolumn{2}{c}{ImageNet-1k} 
    \\\cmidrule(lr){2-3}\cmidrule(lr){4-4}\cmidrule(lr){5-6}
    & \multicolumn{1}{c}{ResNet-9} & \multicolumn{1}{c}{ResNet-50}   & \multicolumn{1}{c}{ResNet-9} 
    & \multicolumn{1}{c}{ResNet-50 @ 128}& \multicolumn{1}{c}{ResNet-50 @ 256} \\\midrule
    ConvBN     & 652 & 2464 & 653 & 1483 & 5891 \\
    GELU       & 45  & 212  & 45  & 341  & 611  \\ 
    Bootstrap  & 189 & 1632 & 188 & 2409 & 4924 \\
    Avg Pool   & 35  & 10   & 35  & 132  & 1114 \\ 
    Linear     & 3   & 3    & 21  & 88   & 469  \\ \midrule
    Total      & $924  \pm 5$ 
               & $4321 \pm 90$ 
               & $942  \pm 5$ 
               & $4453 \pm 83$
               & $13009 \pm  62$\\ \bottomrule
    \end{tabular}
\end{table}


\section{Conclusion and Future Work}
We have successfully constructed three families of ResNet architectures that may be evaluated homomorphically: 1) the multiplexed family of architectures~\cite{lee2022low}, 2) the ResNet-9 bag-of-tricks architectures~\cite{coleman2017dawnbench}, and 3) the popular ResNet-50 architecture~\cite{he2016deep}. Models have been homomorphically evaluated on a variety of standard datasets, including CIFAR-10, CIFAR-100, and ImageNet-1k. We proposed a training time technique to regularize the range of inputs to the GELU activation function by penalizing the fourth order statistical moment of the outputs of the BatchNorm distributions; this technique allows us to efficiently approximate the GELU function with polynomials under homomorphic constraints. When runtimes are compared to the previously reported runtimes of the multiplexed family, we observe a speedup on the previous state-of-the-art by approximately $4.6-6.5\times$ without any classification accuracy degradation. We also report the highest homomorphically encrypted accuracy on CIFAR-10 and ImageNet-1k of $98.3\%$ and $80.2\%$, respectively.

Future work includes extending our models to more advanced tasks, such as encrypted object detection with the YOLO~\cite{redmon2018yolov3} family of architectures and sensitive document analysis with encrypted transformers~\cite{NIPS2017_3f5ee243}. Parallelization in this work was achieved with using multicore CPUs, but vectorized addition and multiplication operations on ciphertexts vectors could be ported to GPUs (or other hardware accelerators) to further accelerate computation and minimize latency.








\bibliographystyle{abbrv}
\bibliography{references.bib}

\section{Appendix}

\subsection{Regularizing Activation Function Inputs With Kurtosis} \label{sect:kurtosis}

Limiting the multiplicative depth of an arithmetic circuit enables faster homomorphic computation, which is of 
particular interest when approximating activation functions with high degree polynomials.
While BatchNorm outputs are designed to be mean-centered 
and normalized by the variance, in practice large asymmetric tails are observed that extend
beyond the desired range, and only gets worse as the model depth increases.

In order to regularize the range of the inputs to the activation function, we have taken an
approach inspired by Ref.~\cite{Shkolnik_2020}, in which the statistical 
moments of hidden layers are taken into account during training. 
Specifically, the original loss function $\mathcal{L}_{\textrm{orig}}$ gets updated to the following:
\begin{align}
    \mathcal{L} = \mathcal{L}_{\textrm{orig}} + \mathcal{L}_{\textrm{BN}},
\end{align}
where $\mathcal{L}_{\textrm{BN}}$ is a loss associated to the BatchNorm distribution, and has
been formulated to incentivize a more Gaussian-shaped distribution that is centered at
zero with a unit standard deviation:
\begin{align}
     \mathcal{L}_{\textrm{BN}} &= \lambda_{\mu} \mathcal{L}_{\mu} + \lambda_\sigma \mathcal{L}_{\sigma} + \lambda_{\kappa} \mathcal{L}_{\kappa}, \\
                               &= \frac{\lambda_{\mu}}{N} \sum_i^N (\mu_i - 0)^2 +
                                  \frac{\lambda_{\sigma}}{N} \sum_i^N (\sigma_i - 1)^2 +
                                  \frac{\lambda_{\kappa}}{N} \sum_i^N (\kappa_i - 3)^2,
\end{align}
where $N$ represents the number of BatchNorm operations performed, and $\mu$, $\sigma$, 
and $\kappa$ are the mean, standard deviation, and kurtosis moments, respectively, of the 
activation function inputs. For this work, our $\lambda$ parameters
are equal and empirically chosen, taking on an epoch dependent value in the range of $[0.0, 0.1]$.

Ref.~\cite{Shkolnik_2020} applied this kurtosis regularization to the convolution weights in order to encourage a uniform boxcar distribution of the kernel elements, as their goal was to enable accurate computation of the convolution using very low-precision fixed point. In contrast, we instead apply kurtosis regularization to the output of the hidden layers to encourage a Gaussian distribution, as our goal is to constrain the range of inputs to the activation function by penalizing large outliers.

\begin{figure}[t]
    \centering
    \includegraphics[scale=0.65]{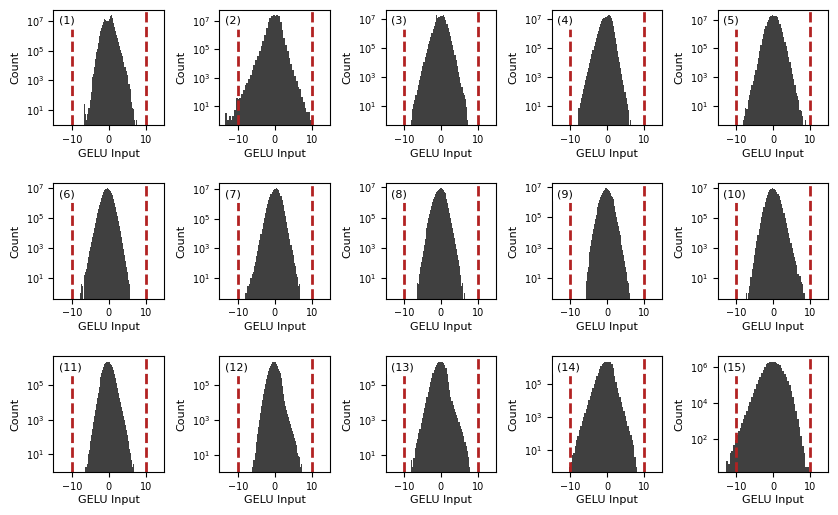}
    \caption{The inputs to all GELU functions for the ResNet-20 architecture and the CIFAR-10 test set. The vertical dotted lines represent a proposed GELU bound.}
    \label{fig:resnet20_gelu}
\end{figure}

Figure~\ref{fig:resnet20_gelu} shows the result of training a ResNet-20 with a $\mathcal{L}_{\textrm{BN}}$ term, where the histograms are the BatchNorm outputs at each layer of the network for the entire CIFAR-10 test set. All of the multiplexed ResNet architectures
have similar pre-activation distributions; however, the ResNet-9 and -50 architectures tend to have a
larger range for the first layer of the network, and then all subsequent layers tend
to fall in the $[-10,10]$ range. This may be attributed to variations in the first layer, for example ResNet-9 uses an untrainable first layer, and ResNet-50 has a large kernel size of 7. Table~\ref{table:GELU_bounds} shows the GELU bound used for all models presented in this work.

\begin{table}
    \centering
    \caption{The chosen GELU bound for all architectures, and the corresponding
    standard deviation in the logit residual, see text for the definition.}
    \label{table:GELU_bounds}
    \begin{tabular}{llcc}
    \toprule
    Dataset   & Model       & GELU Bound & Logit Residual Std.\\ \midrule
    CIFAR-10  & ResNet-9     & 10         & $0.001$ \\ 
              & ResNet-50    & 10         & $0.001$ \\ 
              & ResNet-20*   & 15         & $0.013$ \\
              & ResNet-32*   & 15         & $0.007$ \\
              & ResNet-44*   & 15         & $0.014$ \\
              & ResNet-56*   & 15         & $0.009$ \\
              & ResNet-110*  & 15         & $0.005$ \\ \midrule
    CIFAR-100 & ResNet-9     & 10         & $0.004$ \\
              & ResNet-32*   & 25         & $0.023$ \\ \midrule
    ImageNet-1k & ResNet-50 @ 128  & 15   & $0.001$ \\
    ImageNet-1k & ResNet-50 @ 256  & 15   & $0.001$ \\ \bottomrule
    \end{tabular}
\end{table}


\subsection{Polynomial Approximation}\label{appendix:cheby}


Nonlinear functions are evaluated in RNS-CKKS by using polynomial approximations. Higher degree polynomials require more multiplicative levels, and therefore require either bootstrapping to a higher level or bootstrapping more often, both of which negatively impact performance. For this reason, is it important to limit the degree of the polynomial approximation. This degree depends on several factors including the required accuracy, the particular function being approximated, as well as the input range over which the approximation must be accurate.

Directly evaluating large polynomials in RNS-CKKS can be troublesome, as the coefficients and monomials typically span many orders of magnitude, which is incompatible with RNS-CKKS's noisy fixed point encoding. This problem can be resolved by evaluating the polynomial in a Chebyshev basis. We first review the definition and some properties of the Chebyshev polynomials.

\subsubsection{Chebyshev Basis} \label{sect:poly-approx}

The Chebyshev basis is naturally suited to Chebyshev interpolation, where the approximating polynomial is obtained through Lagrange interpolation of the function at special Chebyshev nodes. However, other approximation techniques such as least squares and minimax optimization can achieve higher accuracy under certain metrics.

Because the ReLU function has a discontinuous derivative at zero, it is difficult to accurately approximate with smooth polynomials. Multiple continuously differentiable variants of ReLU are commonly used in machine learning, including the GELU, Swish, CELU, and Softplus functions. As exhibited in Table~\ref{table:poly-accuracy}, these smoother functions are more amenable to polynomial approximation. For this reason, in this paper we use the GELU activation function in place of ReLU in all of our DCNN. We use the 59-degree polynomial obtained from Chebyshev nodes, as the interpolation was already implemented in OpenFHE and the difference between Chebyshev and Minimax approximation for GELU was negligible. The required depth of 6 is a substantial improvement over the 14 depth approximation relying on composing minimax polynomials used in Ref.~\cite{lee2022low}, which allows us to bootstrap to a lower depth to achieve sufficient depth. Our lower degree ensures that we can use a faster bootstrap operation.

Our polynomial approximation also matches the $\ell_1$-norm accuracy of $< 2^{-13}$ achieved by Ref.~\cite{lee2022low}, albeit for a different function and domain. We emphasize that this improvement is primarily due to our choice of the activation function and the tighter distribution of inputs, rather than in the method for numerical approximation.

\subsubsection{Chebyshev Polynomials} 

The Chebyshev polynomials of the first kind can be recursively defined for even and odd indices by the following properties:
\begin{align}
    T_0(x) &= 1 \\ \label{eq:cheby}
    T_1(x) &= x  \nonumber \\
    &\ldots  \nonumber \\
    T_{2 n}(x) &= 2\; T_n(x)^2 - 1  \nonumber \\
    T_{2 n + 1}(x) &= 2\; T_n(x) \; T_{n+1}(x) - x   \nonumber.
\end{align}
This recursive definition is particularly convenient for RNS-CKKS, as it enables using a binary tree to compute the first $n$ Chebyshev polynomials with $O(n)$ multiplications and a maximum multiplicative depth of $O(\log n)$. This binary tree definition of the Chebyshev polynomials is not common, but it can be derived by applying the following recursive and compositional identities of the Chebyshev polynomials with $m=2$~\cite{kimberling1980four}:
\begin{align*}
T_n(x) &= 2 \;x  \;T_{n-1}(x) - T_{n-2}(x) \;\;\;\forall n \\
T_m(T_n(x)) &= T_{m n} (x) \;\;\;\forall m,n.
\end{align*}

The Chebysev polynomials are commonly used in approximation theory for their numerical stability. They also have the convenient property that they are bounded between $[-1, 1]$ when evaluated over the domain $[-1, 1]$. Furthermore, when used as a basis for polynomial interpolation, the resulting coefficients are typically more bounded than when a basis of monomials is used. Together these properties help avoid the catastrophic cancellation problem that would otherwise arise when homomorphically evaluating high-degree polynomial approximations.

Explicitly, using the Chebyshev basis allows homomorphically evaluating a 59-degree polynomial using a multiplicative depth of 6, or a 27-degree polynomial using a depth of 5.

We note that the extent to which the smoother activation functions can be more accurately approximated than ReLU by fixed degree polynomials depends strongly on both the range over which the approximation must hold as well as the polynomial degree. We examine the specific example of GELU and ReLU with 59 degree minimax approximations, where we use different sets of polynomials for different input ranges. For the wide input range $[-65, 65]$ used in Ref.~\cite{lee2022low}, the GELU approximation has only $1.2\times$ less error than the ReLU approximation. However, for the narrower input range $[-16, 16]$, the GELU approximation has $389\times$ less error than the ReLU approximation! The dramatic difference between these scenarios underscores the impact of tighter input ranges for creating accurate and efficient encrypted DCNNs.

\begin{table}
    \caption{The maximum absolute error of polynomial approximations to the ReLU and GELU activation functions over the domain $[-16, 16]$.}
    \label{table:poly-accuracy}
    \centering
    \begin{tabular}{cccccc}
    \toprule
     Approximation method & degree & ReLU & GELU \\  \midrule 
     Chebyshev nodes & 27 & 0.2862 & 0.0794  \\  
     Minimax & 27 & 0.0866 & 0.0267  \\  
     Chebyshev nodes & 59 & 0.1334 & 0.0002  \\  
     Minimax & 59 & 0.0391 & 0.0001  \\  \bottomrule
    \end{tabular}
\end{table}

\subsection{Precision and Bootstrapping Noise} \label{appendix:metaBTS}

We observed that the error did not noticeably increase for deeper layers in the network. When homomorphically evaluating the ResNet-50 model on CIFAR-10, the average error in each layer remained stable at $0.0002$. This held for all layers, including the final logits. We determined that this error was primarily due to noise from the bootstrapping operation, rather than due to error in the GELU approximation. For this reason, we investigated the impact of the MetaBTS technique~\cite{bae2022meta} for doubling the bootstrap precision.

The MetaBTS technique requires one additional multiplicative level, so in this experiment we used a RNS-CKKS ring dimension of $2^{16}$ and batch size of $2^{15}$ with 59-bit scaling factors and a multiplicative depth of 35. When evaluating the ResNet-50 model on CIFAR-10 with the MetaBTS, the average error in each layer was $0.00001$, resulting in logits with a $20\times$ increased precision. This extremely high precision guarantees that the output of encrypted inference matches that of plaintext inference. The digits of precision increased but did not double after applying the MetaBTS technique, indicating that at this point error from other sources such as in the polynomial approximation became more significant than the reduced bootstrapping noise.
Due to the increased multiplicative depth and number of bootstraps, the latency of inference increased by a factor of $1.7\times$.

As such, the MetaBTS technique appears to be an effective tool for performing very high precision PPML, with a tradeoff between precision and runtime. However, to get the most benefit from the increased bootstrapping precision, other sources of noise including approximation error must also be reduced. Furthermore, using kurtosis-regularized weights in conjunction with GELU activation appears to provide more than sufficient accuracy even without the MetaBTS technique.

\subsection{Training on ImageNet}

We trained from scratch using a particular bag-of-tricks from MosaicML~\cite{mosaicML}, in addition to kurtosis with a fixed hyperparameter $\lambda$ of 0.01. Other hyperparameters were not adjusted or optimized. We expect that minor accuracy gains could be achieved by further tweaking the hyperparameters to account for the architectural changes as well as the additional regularization from kurtosis. Due to the long training time, in the main paper we report the top-1 accuracy for a single training run of each model.

%

\end{document}